\documentclass[runningheads]{llncs}
\usepackage{graphicx}

\usepackage{booktabs}
\usepackage{color}
\usepackage{graphicx}
\usepackage{soul}
\usepackage{tabularx}
\usepackage{url}
\usepackage{verbatim}
\usepackage{xcolor}
\usepackage{subcaption}
\usepackage{hyperref}
\usepackage{multirow}

\usepackage{siunitx}
\newcommand{\acro}[1]{\textsc{#1}}

\begin{document}

\title{The Application of Market-based Multi-Robot Task Allocation to Ambulance Dispatch}  
\titlerunning{Market-based Task Allocation for Ambulance Dispatch}
\author{
Eric Schneider\inst{1} \and 
Marcus Poulton\inst{2} \and 
Archie Drake\inst{1} \and 
Leanne Smith\inst{3} \and \\
George Roussos\inst{2} \and
Simon Parsons\inst{1,4} \and
Elizabeth I Sklar\inst{1,4}
}

\authorrunning{E. Schneider et al.}

\institute{King's College London, WC2R 2LS, UK 
\and
Birkbeck College, University of London, WC1E 7HX, UK
\and
London Ambulance Service, UK\\
\and
University of Lincoln, LN6 7TS, UK\\
\emph{contact author:} \email{esklar@lincoln.ac.uk}
}

\maketitle

\begin{abstract}
Multi-Robot Task Allocation (MRTA) is the problem of distributing a set of tasks to a team of robots with the objective of optimising some criteria, such as minimising the amount of time or energy spent to complete all the tasks or maximising the efficiency of the team's joint activity.
The exploration of MRTA methods is typically restricted to laboratory and field experimentation.
There are few existing real-world models in which teams of autonomous mobile robots are deployed ``in the wild'', e.g., in industrial settings.
In the work presented here, a market-based MRTA approach is applied to the problem of ambulance dispatch, where ambulances are allocated in respond to patients' calls for help.
Ambulances and robots are limited (and perhaps scarce), specialised mobile resources; incidents and tasks represent time-sensitive, specific, potentially unlimited, precisely-located demands for the services which the resources
provide. 
Historical data from the London Ambulance Service describing a set of more than 1 million (anonymised) incidents are used as the basis for evaluating the predicted performance of the market-based approach versus the current, largely manual, method of allocating ambulances to incidents.
Experimental results show statistically significant improvement in response times when using the market-based approach.
\end{abstract}

\section{Introduction}
\label{sec:intro}

The well-studied problem of \emph{multi-robot routing} involves assigning a team of robots to travel to a set of locations such that collisions are avoided and some performance metrics are optimised, such as minimising travel time or distance.
The real-world challenge of emergency vehicle dispatch bears a number of key similarities to multi-robot routing.
\emph{Emergency medical services (EMS)} agencies receive calls (\emph{incidents}) which frequently result in the dispatch of one or more vehicles (\emph{responses}) to the location of the incident.
The determination of which ambulance should respond to which incident is highly complex,
involving consideration of traffic conditions, knowledge of road infrastructure, patient situation and requirements for any specialised healthcare equipment or specially trained personnel.
Sub-optimal travel time or improperly equipped emergency response crews can contribute to loss of life, and extraneous travel distance can result in unnecessary costs for what are typically financially challenged public agencies.

In the study described here, we apply our earlier work on \emph{market-based mechanisms} for \emph{multi-robot task allocation (MRTA)} to the real-world problem of ambulance dispatch.
Conceptually, we map robots to ambulances, tasks to incidents and allocated tasks to ambulance responses.
We present three novel contributions:
\begin{enumerate}
\item The use of historical data from the London Ambulance Service (LAS) that serves as the set of incidents considered---instead of previously engineered or randomly chosen task and robot starting locations, as is the case in most MRTA literature;
\item The use of the next-generation \emph{routing engine} developed for the LAS ---instead of classic routing engines such as A*, which is used by many MRTA approaches, or Google Maps, which is used by many recent studies involving traffic management;
and
\item Experimental evaluation comparing predicted response times when using market-based mechanisms versus the current, largely manual, method of task (incident-to-ambulance) allocation---showing statistically significant differences and improved response times when the market-based mechanisms are utilised.
\end{enumerate}
Although this proof-of-concept study is conducted in collaboration with LAS, and so focuses on London (UK), and on emergency medical services response,  we believe that more general conclusions can be drawn for ambulance services in other parts of the world, as well as other types of emergency response services, such as police or fire.

The structure of this paper is as follows.
Section~\ref{sec:mrteam} covers technical background and prior work on aspects of multi-robot task allocation.
Section~\ref{sec:las} describes the specific application domain addressed in our work: the problem of ambulance dispatch facing the LAS.
Section~\ref{sec:exp} outlines a set of experiments we conducted using data provided by LAS, where we applied market-based multi-robot routing techniques to the ambulance dispatch problem.
Section~\ref{sec:results} presents and discusses the results of these experiments.
Then, Section~\ref{sec:related} highlights related work on ambulance dispatch.
Finally, in Section~\ref{sec:summary}, we conclude with a summary of current work, next steps and directions for future work.

\section{Multi-Robot Task Allocation}
\label{sec:mrteam}

In a multi-robot routing problem, a team of mobile robots must collectively visit some number of \emph{task locations} where they will perform activities such as site inspection or object pick-up and/or delivery.
Solutions to this class of problem entail distributing tasks to robots and
planning routes, or \emph{paths}, from robots' current positions to their assigned task locations in order to optimise some criteria such as travel distance or time.
A multi-robot routing problem is similar to a multiple depot, multiple
Travelling Salesperson Problem (mTSP)~\cite{bektas2006multiple} or Vehicle
Routing Problem (VRP)~\cite{laporte1992vehicle}, where vehicles
need not return to their depots~\cite{lagoudakis:etal:rss05}.
The primary challenge of solving a multi-robot routing problem is \emph{multi-robot
task allocation (MRTA)}: deciding which tasks should be assigned to which
robots so that the overall execution of a mission is, by some measure,
efficient. While there are several kinds of approaches to solving task
allocation problems, we focus on \emph{market-based} methods of task allocation,
and \emph{auctions} in particular, because they can be flexible, distributed and, in some cases, scalable.

Market-based approaches to task allocation frame the task assignment problem as a
multi-agent systems (MAS) problem. Rather than having a centralised
planner be responsible for computing the costs or utilities of potential
allocations, a market-based approach to MRTA relies on the fact that robot team
members are each capable of planning subsets or sub-problems of the mission
(i.e., planning to execute individual tasks or groups of tasks) and can express
the costs or utilities of these plans in a way that is simple and efficient to 
communicate. Task allocation is governed by a \emph{mechanism}: a set of rules
that dictates how tasks should be assigned, and a protocol for communicating
the availability of robots to address tasks,
the availability of tasks for robots to address,
the value robots associate with completed tasks
and
the costs to robots for addressing tasks.
A mechanism enables a virtual marketplace in which tasks can be distributed to
robots or exchanged among them. 
A common kind of market-based mechanism
for MRTA is an \emph{auction}, which compares bids for resources from
interested parties and awards them to the highest (or lowest) bidder according
to the particular rules of a mechanism. It can be expensive to compute an
allocation that is optimal for some performance objective, so most auction
mechanisms strive for \emph{approximately} optimal allocations. Designers of
auction mechanisms must make trade-offs between the costs of computing an
allocation and the performance of the execution of a mission that results from
the allocation.

Our previous work in the MRTA domain led to the development of our \acro{MRTeAm} framework~\cite{schneider:thesis,schneider-et-al-taros:2015}, which was designed to evaluate a range of task allocation mechanisms both in simulation and on physical robots.
While other research in multi-robot routing has concentrated largely on discovering optimal assignment mechanisms for a single type of environment, our work using \acro{MRTeAm} has focussed on evaluating a range of performance metrics in a variety of complex task environments~\cite{gerkey2004formal,landen2010complex} and analysing both task assignment \emph{and} task execution---which makes this framework particularly relevant for application to a real-world domain where we are especially concerned with measuring \emph{response} or \emph{travel times}.

\begin{figure}[t!]
\begin{center}
  \includegraphics[width=0.6\textwidth]{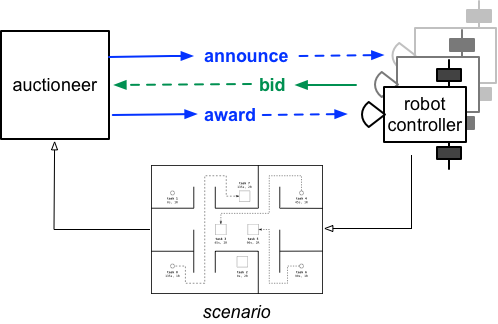}
\end{center}
  \caption{Interaction of an \emph{auctioneer agent} with robots in \acro{MRTeAm}. \emph{Robot controller agents} compute bids and are responsible for autonomous navigation to task locations after allocation. In the ambulance domain, the auctioneer is a proxy for a dispatcher and robot controllers are proxies for ambulance vehicles.}
  \label{fig:mrteam-agents}
\end{figure}

In \acro{MRTeAm}, a \emph{map} specifies the extent of a geographical space and
the arrangements of free space and obstacles within it. A \emph{team}
is a set of $n$ robots $R = \{r_0, \dots, r_{n-1}\}$. 
A \emph{starting configuration}, $S$, specifies the location 
of each robot on the map 
at the beginning of a \emph{mission}.
A \emph{scenario} is a set of $m$ tasks $T = \{t_0,
\dots, t_{m-1}\}$ situated on the map.  Each task $t \in T$ has the
following properties: $t.pos$, a fixed position on the map; $t.arr$,
the arrival time of the task; and $t.req$, the number of robots
required to complete the task.
A \emph{mission} comprises the map, a scenario, and a robot team with a starting
configuration: $M = \{map,\;T,\;R,\;S\}$. An auction mechanism allocates
tasks to robots over a number of \emph{rounds}. In an auction round, a
coordinating \emph{auctioneer agent} announces tasks to the team, team members
compute and submit \emph{bids} to the auctioneer, and the auctioneer awards one or
more tasks to team members according to the rules of the mechanism (Figure~\ref{fig:mrteam-agents}).

A bid for a task is computed by a robot as:
\[
\mathit{bid} = w_0 f_0 + w_1 f_1 + \ldots + w_{\varphi-1} f_{\varphi-1}
\]
where 
$f_i$ is a quantitative factor,
$w_i$ is a weight associated with that factor,
and
$\varphi$ is the number of factors to consider in the bid.
For example, $f_0$ could be the estimated travel distance from the robot's
starting location at the time of placing the bid to the location of the task
it is bidding on, and  $f_1$ might be the priority of the task. If we deem
distance more important than task priority, then $w_0 > w_1$. 

The metrics we use to evaluate performance in \acro{MRTeAm} measure the \emph{distance travelled} by robots as well as various time-based performance measures
such as:
\emph{deliberation time} (the time taken to compute a set of task assignments);
\emph{execution time} (the time taken to execute the assigned tasks);
\emph{movement time} (the time robots spend actually moving towards task locations);
\emph{idle time} (the time a robot sits idly waiting for other robots to complete their tasks after it has completed its last
task);
and
\emph{delay time} (the time each robot spends waiting for other robots to pass safely in order to avoid collisions).
As described in Section~\ref{sec:exp}, \emph{movement time} is the most relevant measure for the experiments described here, though \emph{idle time} will be a key metric to consider in future work.

The next section explains how the problems faced in ambulance dispatch are related to those explored in the \acro{MRTeAm} project.

\section{Ambulance dispatch in London}
\label{sec:las}

\begin{figure}[h!]
\begin{center}
\includegraphics[width=\textwidth]{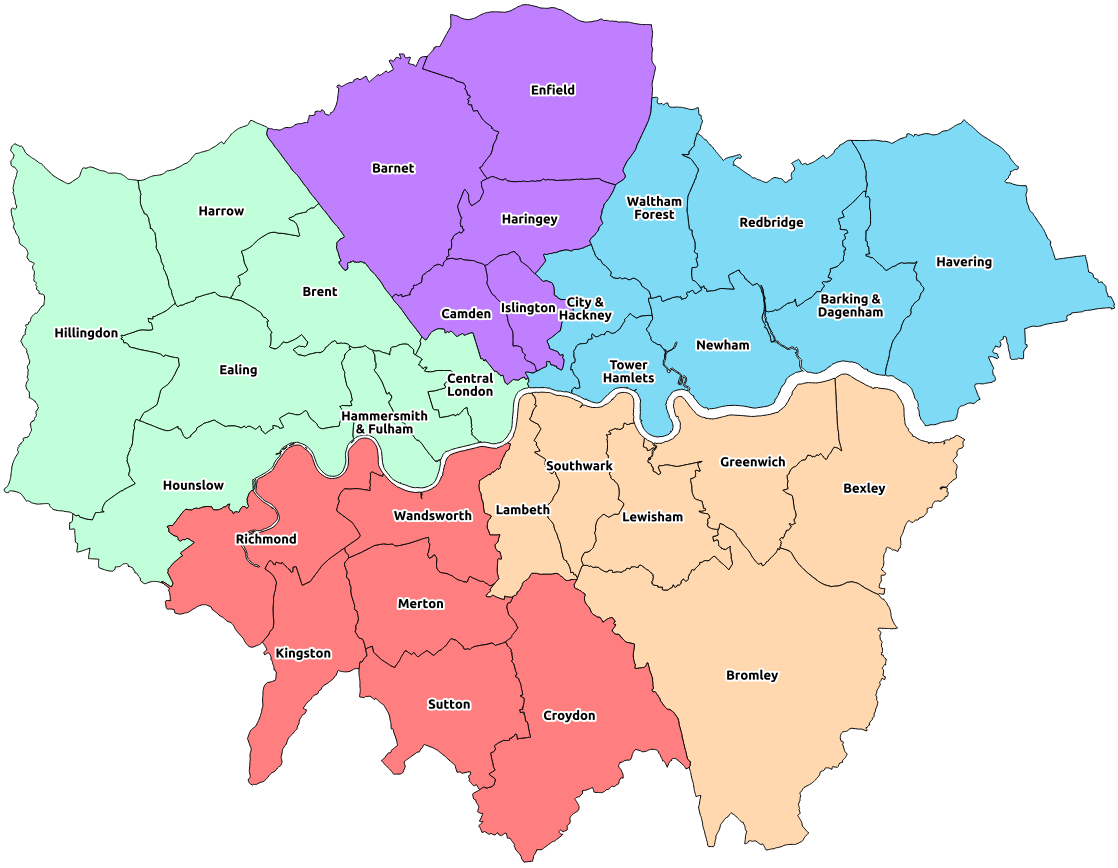}
\end{center}
  \caption{London Ambulance Service sectors and NHS Clinical Commissioning Groups. The Service sectors are the large coloured regions: North Central (purple), North East (blue), North West (green), South East (ochre) and South West (red).
  Clinical Commissioning Groups are the areas within the service sectors.}
  \label{fig:ccgs-sectors}
\end{figure}

Greater London (UK) has a population of approximately 8.2 million people, according to the 2011 national census~\cite{census2011uk},
and its medical emergencies are handled by the London
Ambulance Service, which is governed by a UK state agency, the 
National Health Service (NHS).
In order to manage the needs and resources of such a large city, the NHS
divides London into 33 \emph{Clinical Commissioning Groups} or \emph{CCGs}.
These CCGs are grouped into five \emph{sectors} (Figure~\ref{fig:ccgs-sectors}).
Ambulances are allocated to ``home'' locations within each CCG, indicating where
crews report at the start and end of a shift. In performing assignment of an
ambulance to an incident, there is an attempt to keep ambulances within their
``home'' CCGs for a number of reasons, such as reducing travel distances (and
associated petrol costs) and travel times (not only because distances are
shorter but also because crews are more familiar with the roadways in their home
CCG).

LAS receives emergency medical service calls 24 hours a day, 365 days of the year.
For example, in
2016, LAS handled approximately 5,000 emergency calls on a daily basis. If a call
requires a responder vehicle, then the call is logged as an \emph{incident}.
The main responder vehicles are \emph{accident and emergency units}
(AEUs), the large ``truck'' ambulances which are capable of transporting
a patient to hospital, and the \emph{fast response units} (FRUs), i.e.,
estate cars, which can get to the scene in a shorter amount of time. Every responder
vehicle that is sent to an incident is defined as a separate \emph{response}, so often
there can be multiple responses for a given incident. It is not
unusual for an FRU to be dispatched first, 
in order to give help as quickly as possible,
followed by an AEU, to provide the means to convey a
patient to hospital.

LAS receives many different types of call, and these are categorised based
on severity, which is defined according to the nature of a patient's
\emph{chief complaint}.
Starting in 2017, the LAS adopted new categories 
of severity and performance measurement through the \emph{Ambulance Response
Program (ARP)}\footnote{\url{https://www.england.nhs.uk/urgent-emergency-care/arp/}}.
The work presented here is based on a data set from 2016, so the definitions that follow are from the pre-ARP period.
Calls were categorised from {\sf A} (highest priority) to {\sf C},
each with subcategories referring to the target response time.
For Category {\sf A},
the highest subcategory was \texttt{red1}, referring to a life threatening
incident with a target response time of 8 minutes. The subcategories for
Category {\sf C} ranged from \texttt{green1}, with a target response time of 20 minutes to
\texttt{green4} with a target time of 4 hours.
Depending on the nature of the call, the LAS definition of 
different time measures varies. 
\emph{Response time} is always defined as the difference in time between the \emph{clock start
time} and the first responder arriving at the scene of an incident, however the clock start time is
measured differently for different patients. The clock start for the highest
priority cases begins when the call is answered by the control room. For other
calls, the clock start time begins at the earliest of: the first vehicle being
dispatched; the type of incident being determined; or 240 seconds after the call
is answered.

Performance is measured by the proportion of first-responder response times
that fall within an incident category's maximum allowable time. Performance
is measured for Greater London overall, but also for each CCG. In 2016,
approximately 65\% of Category {\sf A} incidents received responses within
the target 8-minute time limit.

The notion of computer-assisted dispatch (CAD) was first introduced in the London Ambulance Service in 1992 and quickly became a lesson in software engineering mishaps.
The early version of CAD included two key components: ``an
automated vehicle locating system (AVLS) and mobile data terminals (MDTs) to support automatic communication with
ambulances''~\cite{finkelstein-dowell:1996}. Within hours after deployment, the
AVLS lost track of vehicles' whereabouts, so the CAD database
became inaccurate. Ambulances were dispatched non-optimally: some calls received
multiple ambulances; others received none. The CAD software started
issuing error messages and overloaded the system. Ambulance crews stopped
sending status reports via MDTs because the system was too slow. This 
catastrophic failure led to deficient patient care, possible loss of life, and
loss of employment for the LAS chief executive.
Since then, the road has not been smooth: a software upgrade in 2006 led to
systemic failure~\cite{long-wired:2009} and the initial introduction in 2011 of the current CAD system, 
\emph{CommandPoint},\footnote{\url{https://www.northropgrummaninternational.com/capabilities/command-point/}}
was delayed for technical reasons.
However, since 2012, CommandPoint has been successfully providing dispatch support for LAS~\cite{hscic:2013}.
In 2018, the LAS began a process of re-evaluating and upgrading their
CAD software modules to exploit new and emerging technologies,
such as pervasive mobile computing, and sources of data such as real-time traffic
and weather information~\cite{dash2018policy}.

The long term vision for the work described here is the integration of our auction-based multi-robot routing methodology into a computer-assisted dispatch system.
The experiments described in the next section and results that follow will help us demonstrate 
the predicted advantages of our approach.

\section{Experiments}
\label{sec:exp}

This section describes the series of experiments that we conducted to compare the results when vehicles are allocated to incidents using our auction-based mechanism versus the manual allocation process currently employed, where human dispatchers in the LAS control room can consult the CAD system for recommendations but ultimately perform task (vehicle to incident) assignment themselves (manually).

\subsection{Experimental Setup}
The auction-based mechanism employed in our experiments was taken from the \acro{MRTeAm} framework, described earlier.
For simplicity, bids were derived using one quantitative factor: $f_0=\mathit{estimated~travel~time}$.
This provides us with a baseline for future work in which we can consider additional factors in the bidding.
The experiments conducted here demonstrate that even using just one factor, the auction-based methodology predicts significantly shorter response times (detailed in Section~\ref{sec:results}).

Our experimental evaluation was facilitated by an historical data set provided by the LAS, which records, for
each incident that occurred in 2016, the location and call time of the
incident, the locations of vehicles at the times they were dispatched to the
incident, and the vehicles' travel times to the incident location, as well as other information about the incident, such as chief complaint and category.
The data set contains 1.1 million incident records and 1.5 million response records.
In order to keep the data anonymised,
location coordinates are quantised to the nearest vertex on a 100m-precision
grid\footnote{\url{https://www.ordnancesurvey.co.uk/resources/maps-and-geographic-resources/the-national-grid.html}}.
For the experiments described here, we only considered Category {\sf A} incidents.

Because estimated travel time is taken as the basis for bidding, it is important for us to compute that carefully.
We considered two methods for computing routes and estimating travel times between vehicle and incident locations:
one makes use of a publicly available route planner (the Google Maps Directions API,\footnote{\url{https://developers.google.com/maps/documentation/directions}} referred to here as \acro{GMaps}),
and
the other makes use of a proprietary routing engine, called \acro{Quest}~\cite{poulton2013towards}.
Thus we can compare three different response times:
(i)~the \emph{historically observed} response time (taken directly from the LAS data set);
(ii)~the \acro{Quest}\emph{-simulated} response time, taking the vehicle start and end locations from the LAS data set and using the \acro{Quest} routing engine to estimate travel time;
and
(iii)~the \acro{GMaps}\emph{-simulated} response time, again taking the vehicle start and end locations from the LAS data set, but using the Google Maps route planner to estimate travel time.
For privacy reasons, we do not have access to the actual routes taken by emergency response vehicles, so using the \acro{Quest}-simulated routing and response times based on historical start and end locations gives us a fair basis for comparison between actual vehicle choices and hypothetical choices taken by our auction-based mechanism.

We designed a set of simulation experiments to compare two independent variables:
(1)~\emph{vehicle selection} (``historical choice'' (\acro{hist}) or ``auction mechanism choice'' (\acro{auct}))
and
(2)~\emph{routing engine} (\acro{Quest} or \acro{GMaps}).
\textbf{Our hypothesis is that the auction mechanism choices will predict shorter response times than the historical choices, for either routing engine.}
We evaluate this hypothesis in two steps.
First, we produce a benchmark measure by comparing \acro{Quest}-simulated and \acro{GMaps}-simulated travel times with historically observed travel times, using the same (historic) start and end locations from the LAS data set for all three metrics.
Second, we evaluate the efficacy of the auction mechanism by comparing simulated travel times for pairs of start and end locations: the historically recorded vehicle in the LAS data set (the benchmark) versus the vehicle chosen by the auction mechanism.
Details of the two steps are provided below.

\subsection{Benchmark Generation}
To provide a benchmark for evaluation, we compared the historically observed response time for a chosen vehicle from the LAS data set with the simulated response times computed by each routing engine.
In both cases, we used the start and end locations of the historically chosen vehicles to compute routes.
This step also serves to demonstrate the advantage of using \acro{Quest}-simulated travel times, which are derived from historical road speeds of emergency service vehicles driving along road segments in London and includes travel permitted only by blue light (emergency) vehicles.
A sample of 2000 Category {\sf A} incidents was drawn
uniformly randomly from the data set. We identified the first response vehicle
assigned to each incident and queried each of \acro{Quest} and \acro{GMaps} for a route
between the vehicle's location at the time the vehicle was dispatched and the incident's
location, along with an estimated travel time. We then compared both estimates
to the historical travel time observed in the sample. The results, shown in Figure
\ref{fig:response-distributions} and discussed in the next section, show that the estimates computed by \acro{Quest} are more accurate than \acro{GMaps} with respect to actual travel times of emergency vehicles. Following this demonstration of the effectiveness of the \acro{Quest} engine we carried out the experiments that are the main contribution
of this paper.

\subsection{Adaptation of Auction Mechanism Framework}
The \acro{MRTeAm} framework was adapted for these experiments in the following ways. The \emph{map},
represented in the \acro{Quest} routing engine, is based on the ITN Road Layer\footnote{\url{https://www.ordnancesurvey.co.uk/business-and-government/help-and-support/products/itn-layer.html}} map produced by the UK's Ordnance Survey mapping agency.
Each ``robot'' in $R$ represents an emergency vehicle (ambulance) and is
capable of planning a route between two locations on the map and computing
the distance and estimated time to travel along the route. A \emph{scenario}
comprises a single emergency incident task, described in detail below. The
\emph{mechanism} employed in all experiments is the \emph{sequential single-item}
auction, which has been shown to produce allocations that are close to optimal~\cite{koenig2006} while scaling better than e.g., combinatorial
auctions~\cite{berhault2003robot}. A \emph{bid} comprises a single bid factor,
the estimated travel time between a vehicle's location and an incident location.
The \emph{auctioneer} agent functions identically to its robot setting.

\begin{figure}[t!]
	\centering
	\includegraphics[width=0.6\textwidth]{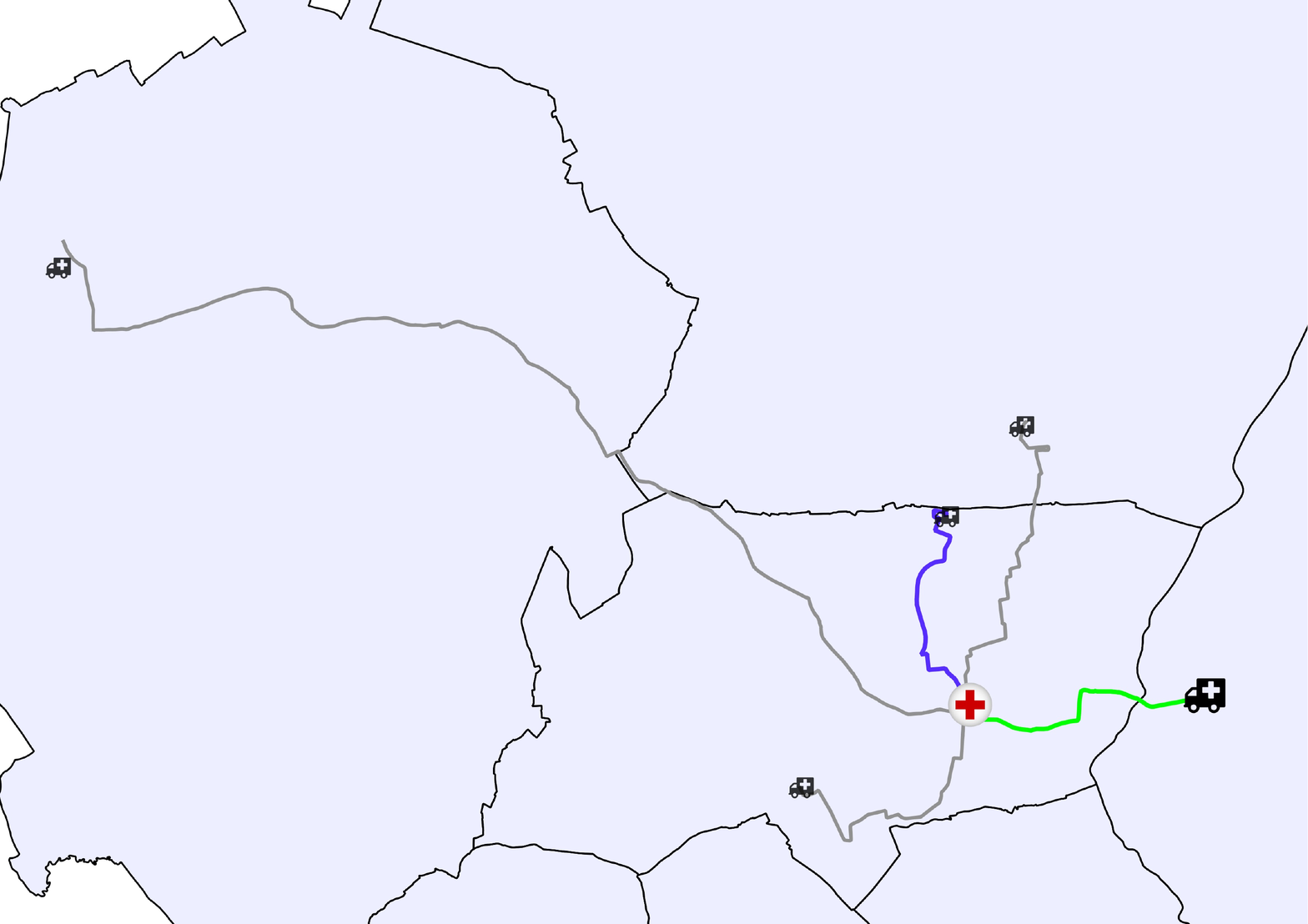}\\[0.1cm]
	\caption[Auction routes example]{Locations of idle vehicles and their routes to an example incident location (red cross). The route (369 seconds) of the historically chosen first-responder vehicle dispatched to the incident is shown in blue. The route of a first-responder vehicle hypothetically chosen via auction (duration: 272 seconds) is shown in green.}
	\label{fig:auction-routes-example}
\end{figure}

The data set did not provide the locations of \emph{idle vehicles}---vehicles
not \emph{en route} to an incident or otherwise assigned (for example, for conveyance
from an incident location to a hospital). However, the data set did provide the
locations of vehicles both at the time they completed their assignments and at the
time they were next dispatched following a completed assignment. Thus, the
locations of idle vehicles for a given time instant were interpolated along a
route between these two locations.

For every incident in each experimental condition, idle vehicles were identified
within a 20 km$^2$ neighbourhood around the incident and their locations at the call time of the incident were estimated as described above.
The auctioneer agent announced the ``task'' (incident) to agents representing idle vehicles in their neighbourhood.
These agents computed bids representing their estimated travel times to the incident location and submitted them to the auctioneer. The auctioneer
then aggregated the bids received and assigned the lowest-bidding vehicle to the
incident. Figure \ref{fig:auction-routes-example} depicts an example in which
idle vehicles have planned routes to an incident location (red cross). The
route of the lowest-bidding vehicle (hypothetically chosen by auction) is shown in green while that of the vehicle actually (historically) dispatched to the incident is shown in blue.

\subsection{Experimental Conditions}
Since evaluating a dispatch decision for every incident from the data set
(\hbox{$>1$} million) was infeasible, we defined four experimental conditions that drew
samples of incidents. In each condition, 100 Category {\sf A} incidents
were sampled uniformly randomly from a temporal and geographic range. For each
sample incident, the travel time 
of the historically-assigned first-responding vehicle from its location at dispatch time to the incident location was compared to that of a (possibly different) vehicle chosen by the auction mechanism.
Condition {\sf 1M-1C} sampled incidents that occurred over one month in one arbitrarily selected Clinical Commissioning Group (CCG); 
{\sf 12M-1C} sampled from 12 months (all of 2016) in the
same CCG;
{\sf 1M-nC} sampled from one month and all CCGs ($n=33$); and
{\sf 12M-nC}
sampled from 12 months and all CCGs. Table~\ref{tab:exp-cond} lists the four
conditions under which experiments were conducted in order to evaluate the
effectiveness of our approach.

\begin{table}[h!]
\begin{center}
\begingroup
\renewcommand*{\arraystretch}{1.5}
\begin{tabular}{r||c|c|}
             & one month & 12 months \\
\hline
\hline
one CCG  & {\sf 1M-1C} & {\sf 12M-1C} \\
\hline
all CCGs & {\sf 1M-nC} & {\sf 12M-nC} \\
\hline
\end{tabular}
\endgroup
\caption{Experimental conditions}
\label{tab:exp-cond}
\end{center}
\end{table}

\subsection{Metrics}
We computed two types of metrics, both of which are analysed in the next section.
The first metric is \emph{simulated response time}, discussed above, where a routing engine takes as input a start and end location for a vehicle and then estimates the amount of time needed for the vehicle to travel from one location to the other.
Shorter response times are better.
This metric is the equivalent to \emph{movement time} from the \acro{MRTeAm} framework.
The second metric is \emph{vehicle choice}.
During experiments, we record the identity of the vehicle chosen by the auction mechanism and then compare that to the historically observed vehicle choice.
We count how many choices were made differently by the auction mechanism as opposed to the human-in-the-loop CAD-advised process currently employed in the LAS control room.
We express these as percentages: higher values indicate more differences in vehicle choice.

\section{Results}
\label{sec:results}

\subsection{Benchmark Generation Results}
\begin{figure}[t!]
	\centering
	\includegraphics[width=0.6\columnwidth]{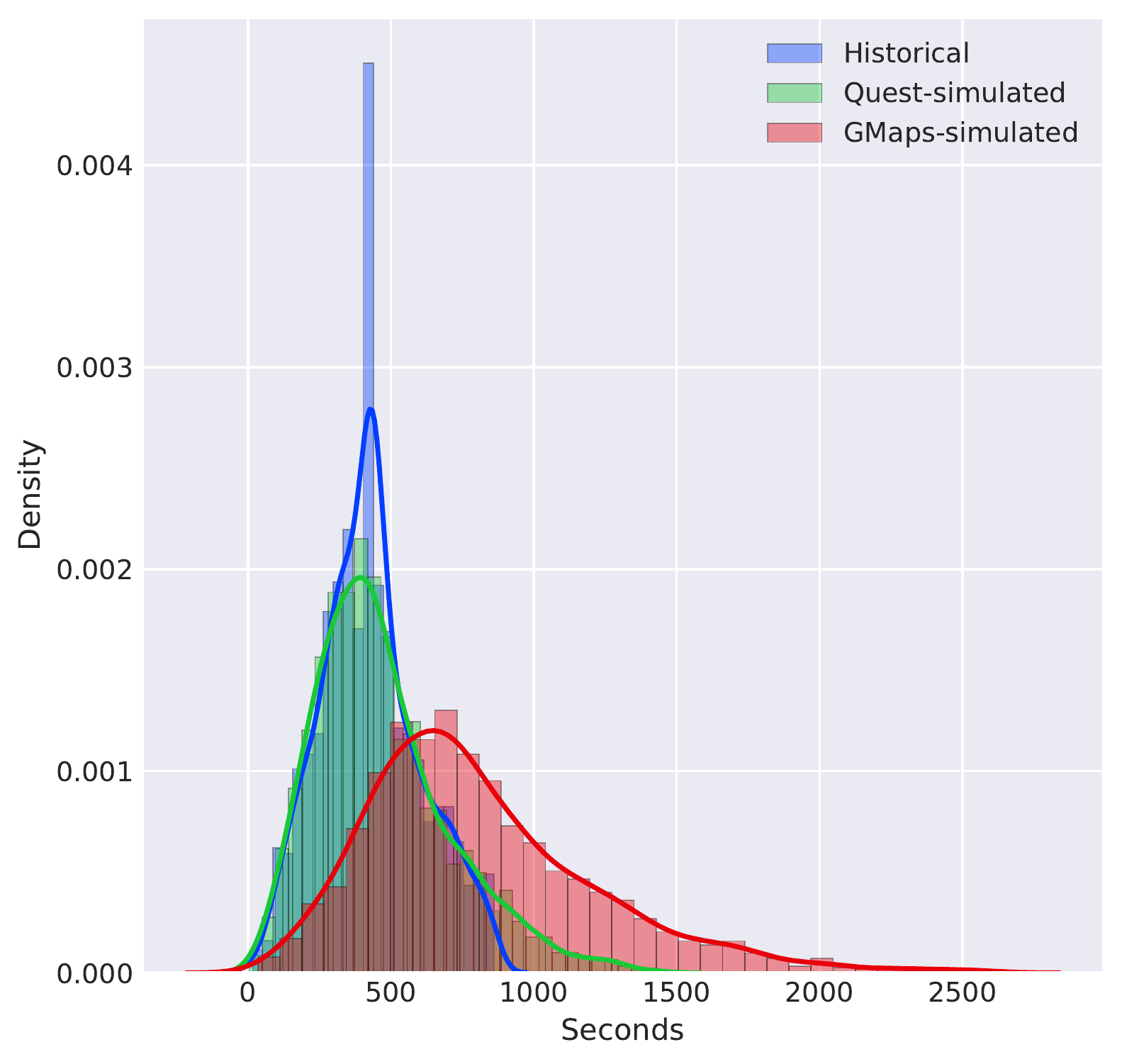}\\[0.1cm]
	\caption[Travel time distributions]{Distributions of travel times to incidents (seconds). Historical travel times (blue) compare with those of the same journeys estimated by the Quest routing engine (green) and the Google Maps Directions API (red).}
	\label{fig:response-distributions}
\end{figure}

Figure \ref{fig:response-distributions} compares distributions of travel
times for 2000 journeys between vehicle and incident locations. Historical travel
times are shown in blue ($\mu=426$s); travel times for the same journeys estimated
by \acro{quest} are shown in green ($\mu=441$s) and those estimated by
\acro{gmaps} are shown in red ($\mu=768$s). The Wasserstein distance~\cite{wasserstein-distance}
from the historically observed distribution of journey times is 40.9 for
\acro{quest} and 336.39 for \acro{gmaps}. 
These results show that \acro{quest} produces travel time estimates that 
closely agree with historical travel times while \acro{gmaps} tends to overestimate them.
Both routing engines employ traffic models that are tuned for specific journey
times (a given hour- or minute-of-the-week). However, \acro{quest}'s estimates are based
on road speeds of emergency vehicles, which obey different traffic rules and
tend to be higher than those of passenger or commercial vehicles, which \acro{gmaps}
targets. These results validate and extend previous work that demonstrated the
accuracy of the \acro{quest} routing engine when compared to a simple model of
computing travel times based on straight-line Euclidean distances 
\cite{poulton2013towards}. The results presented here demonstrate
\acro{quest}'s accuracy even when compared with \acro{gmaps}, a state-of-the-art
routing engine. These benchmark results also provide a measure of confidence in 
the accuracy of our auction-based results as compared with related work
that employed \acro{gmaps} \cite{lujak2013coordinating}.

\subsection{Auction Results}
\begin{figure*}
\begin{center}
  \begin{subfigure}[b]{0.45\columnwidth}
    \includegraphics[width=\textwidth,height=0.2\textheight]{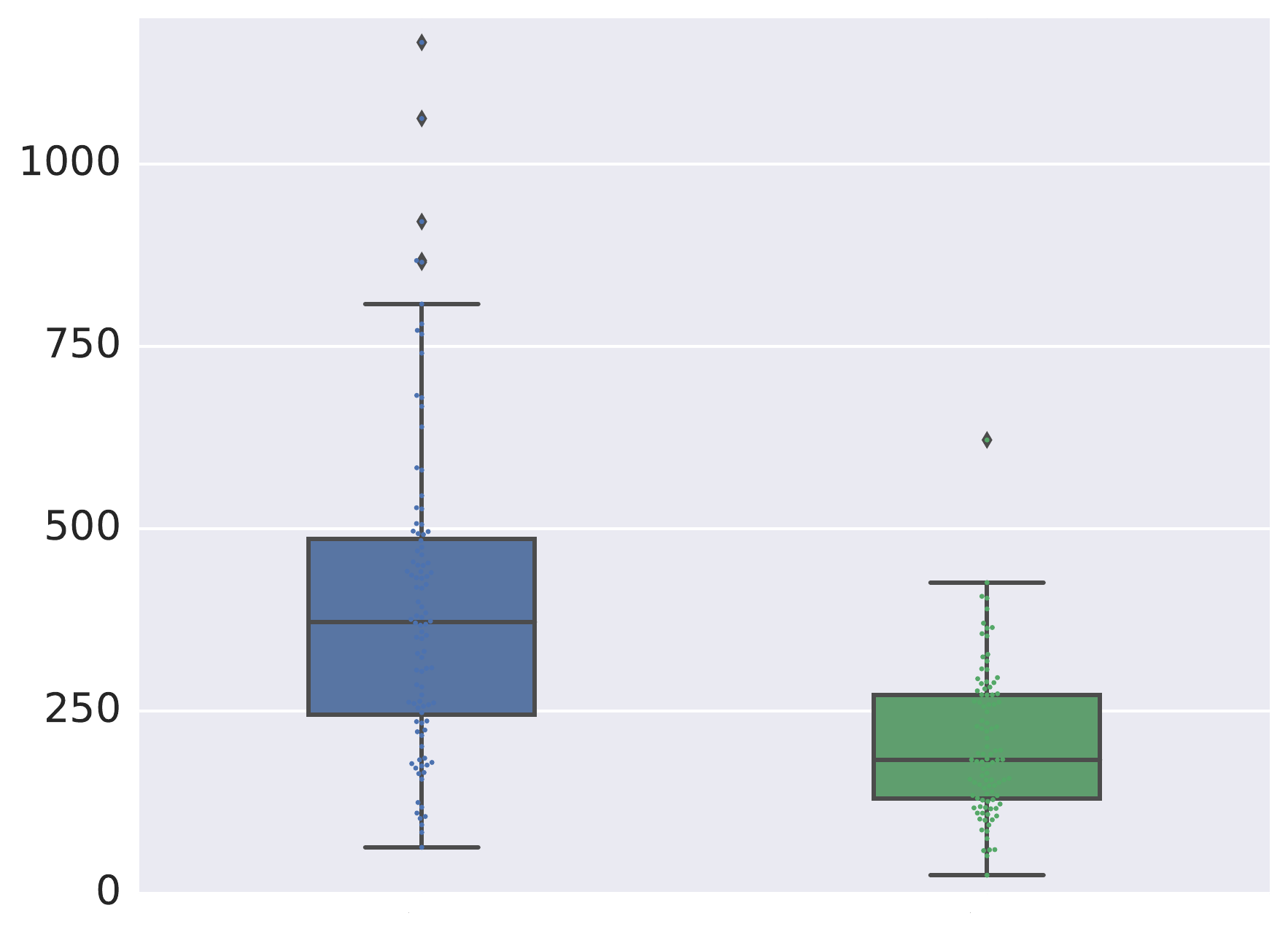}
    \caption[1M-1C Box plot]{1M-1C Box plot}
    \label{fig:exp:1m-1c-box}
  \end{subfigure}
  \begin{subfigure}[b]{0.45\columnwidth}
    \includegraphics[width=\textwidth,height=0.2\textheight]{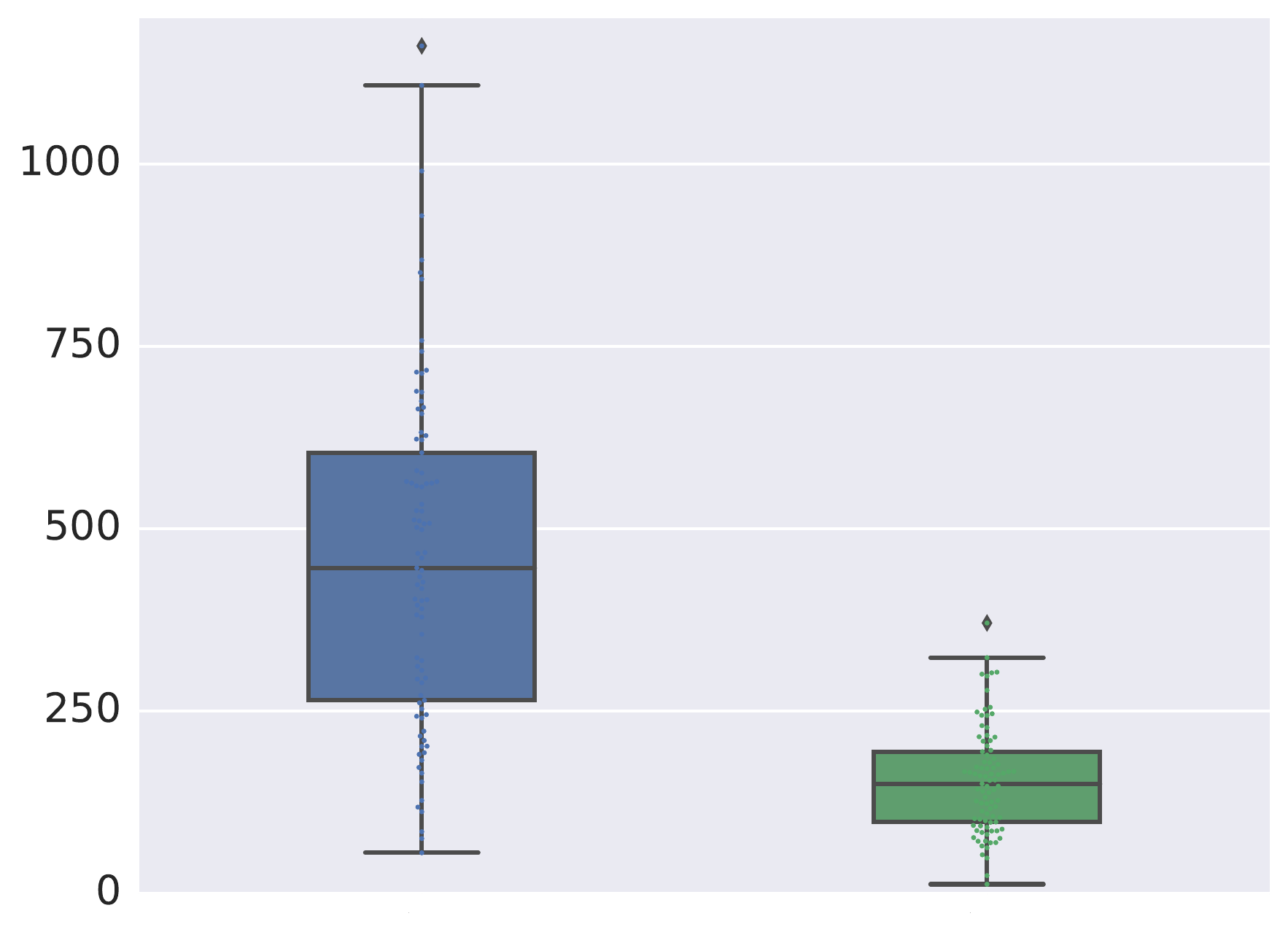}
	\caption[12M-1C Box plot]{12M-1C Box plot}
	\label{fig:exp:12m-1c-box}
  \end{subfigure}\\
  \begin{subfigure}[b]{0.45\columnwidth}
    \includegraphics[width=\textwidth,height=0.2\textheight]{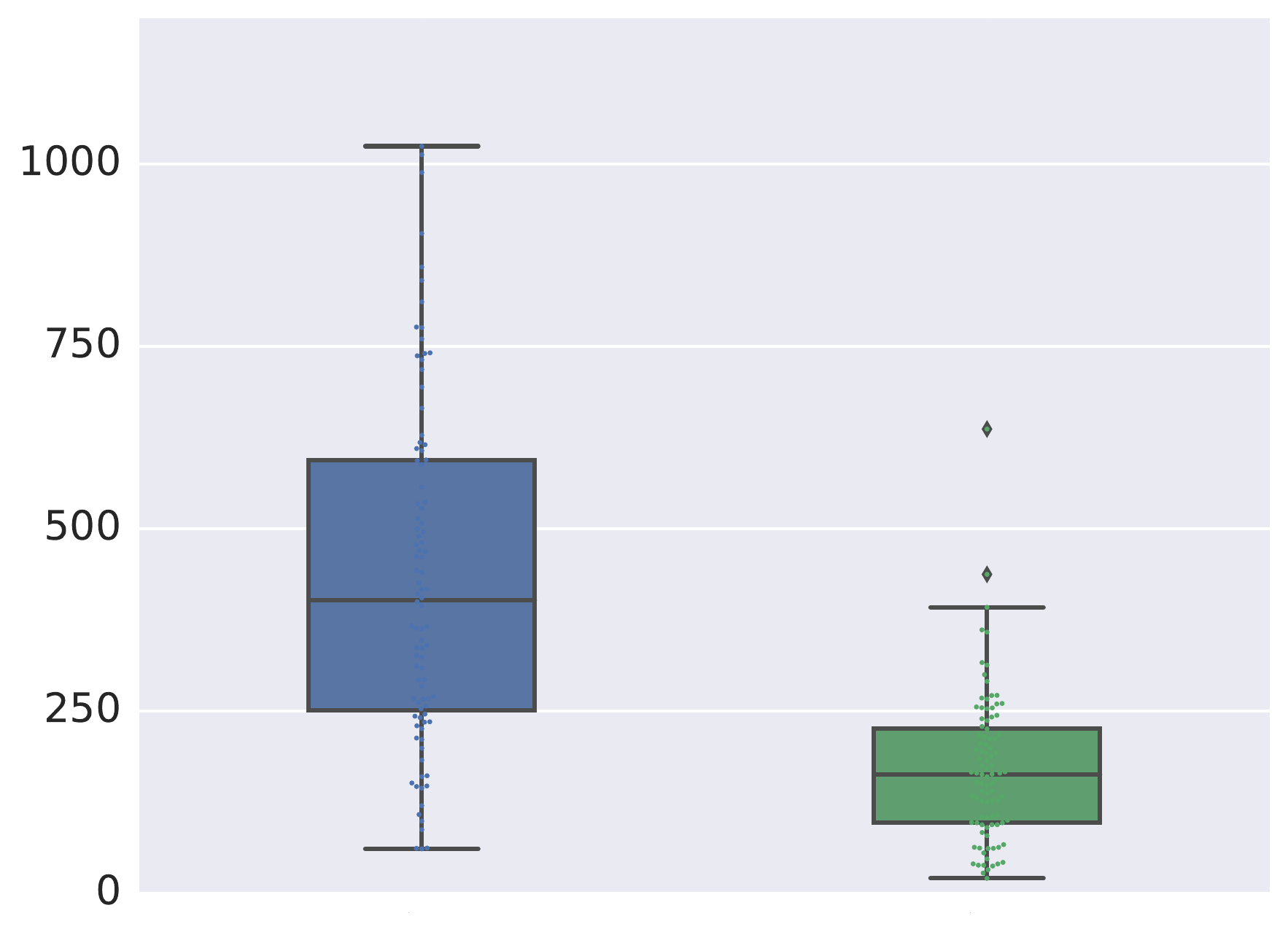}
	\caption[1M-nC Box plot]{1M-nC Box plot}
	\label{fig:exp:1m-nc-box}
  \end{subfigure}
  \begin{subfigure}[b]{0.45\columnwidth}
    \includegraphics[width=\textwidth,height=0.2\textheight]{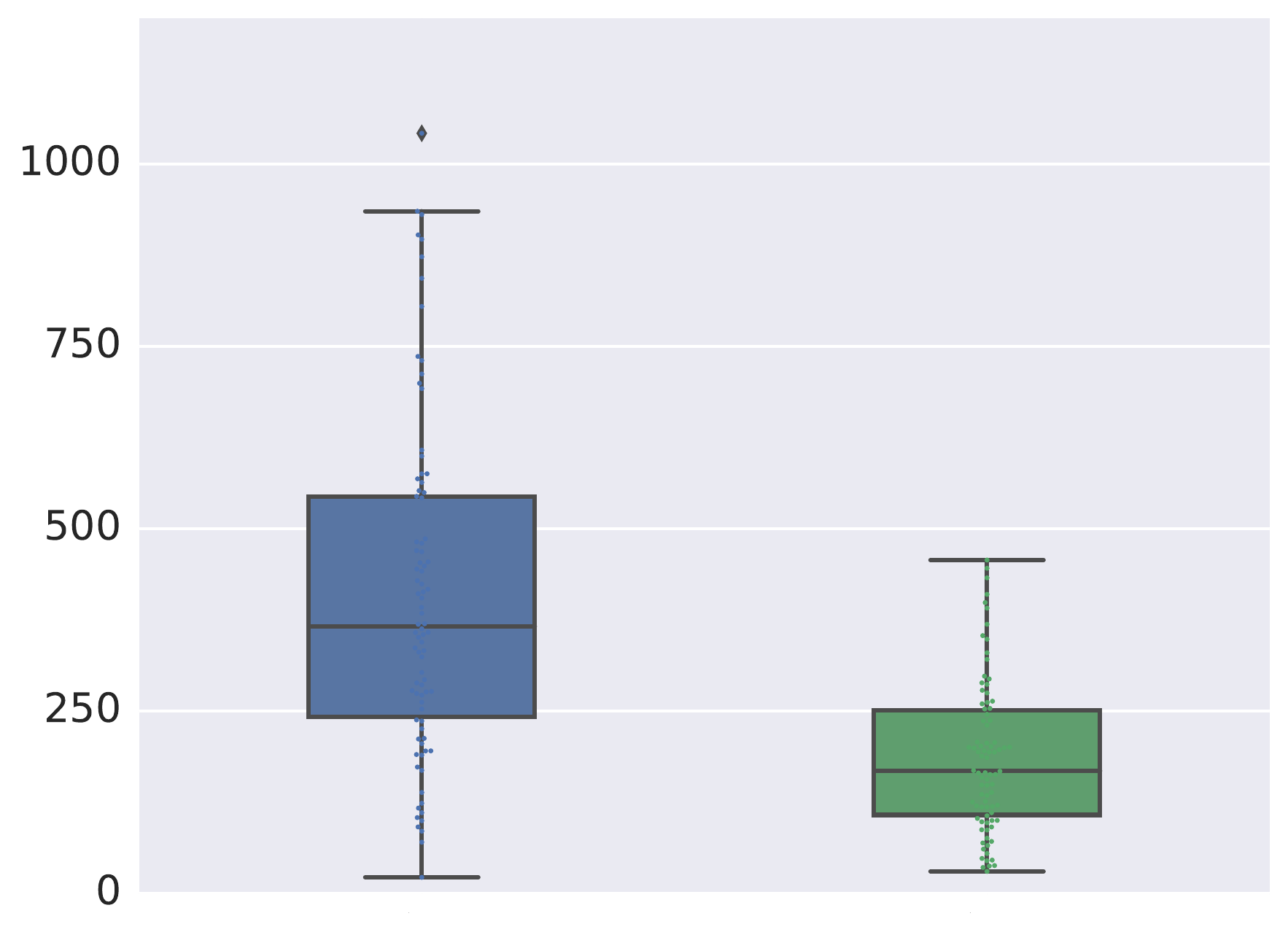}
	\caption[12M-nC Box plot]{12M-nC Box plot}
	\label{fig:exp:12m-nc-box}
  \end{subfigure}\\
  \begin{subfigure}[b]{0.45\columnwidth}
    \includegraphics[width=\textwidth]{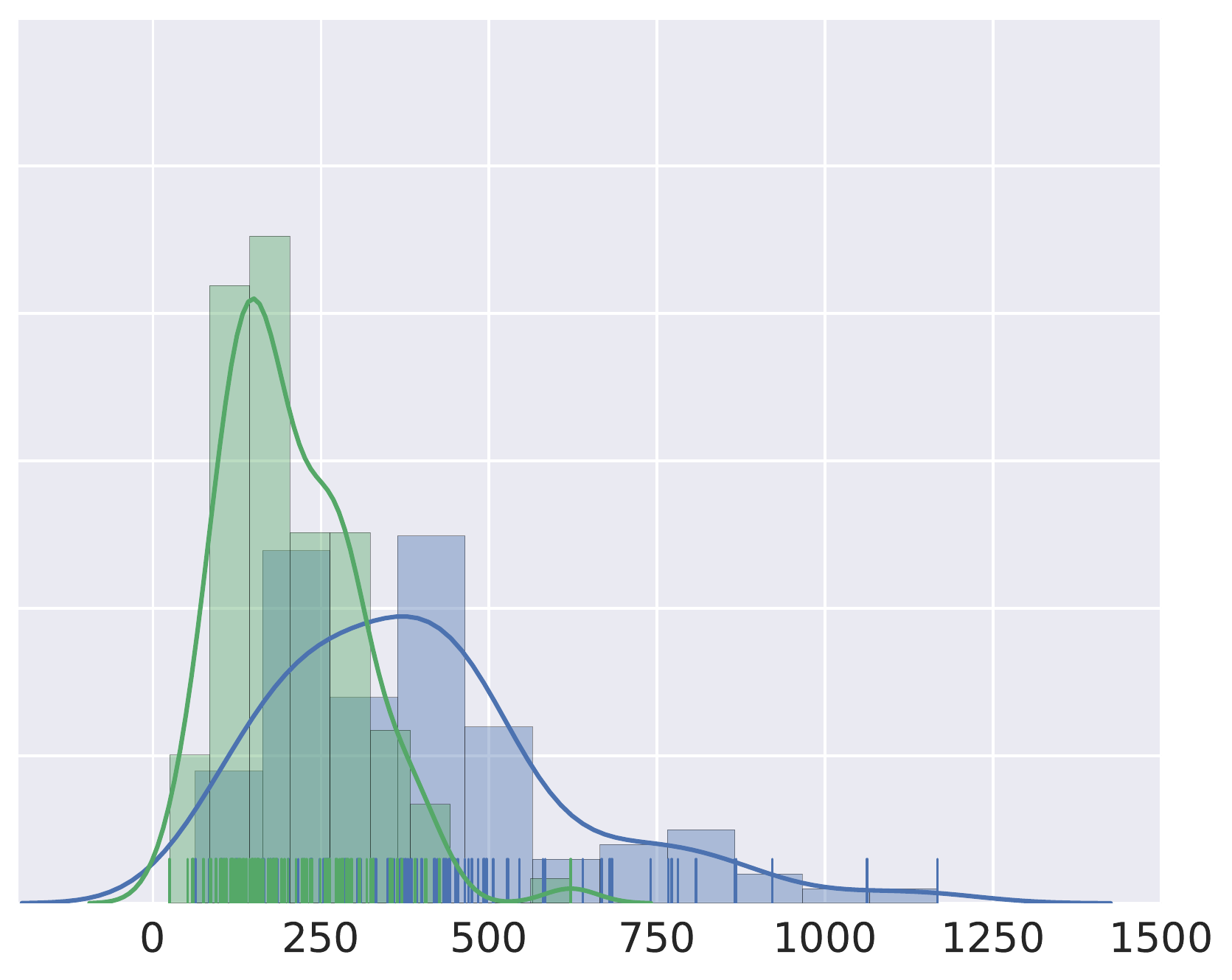}
	\caption[1M-1C distribution plot]{1M-1C Distribution}
	\label{fig:exp:1m-1c-distribution}
  \end{subfigure}
  \begin{subfigure}[b]{0.45\columnwidth}
    \includegraphics[width=\textwidth]{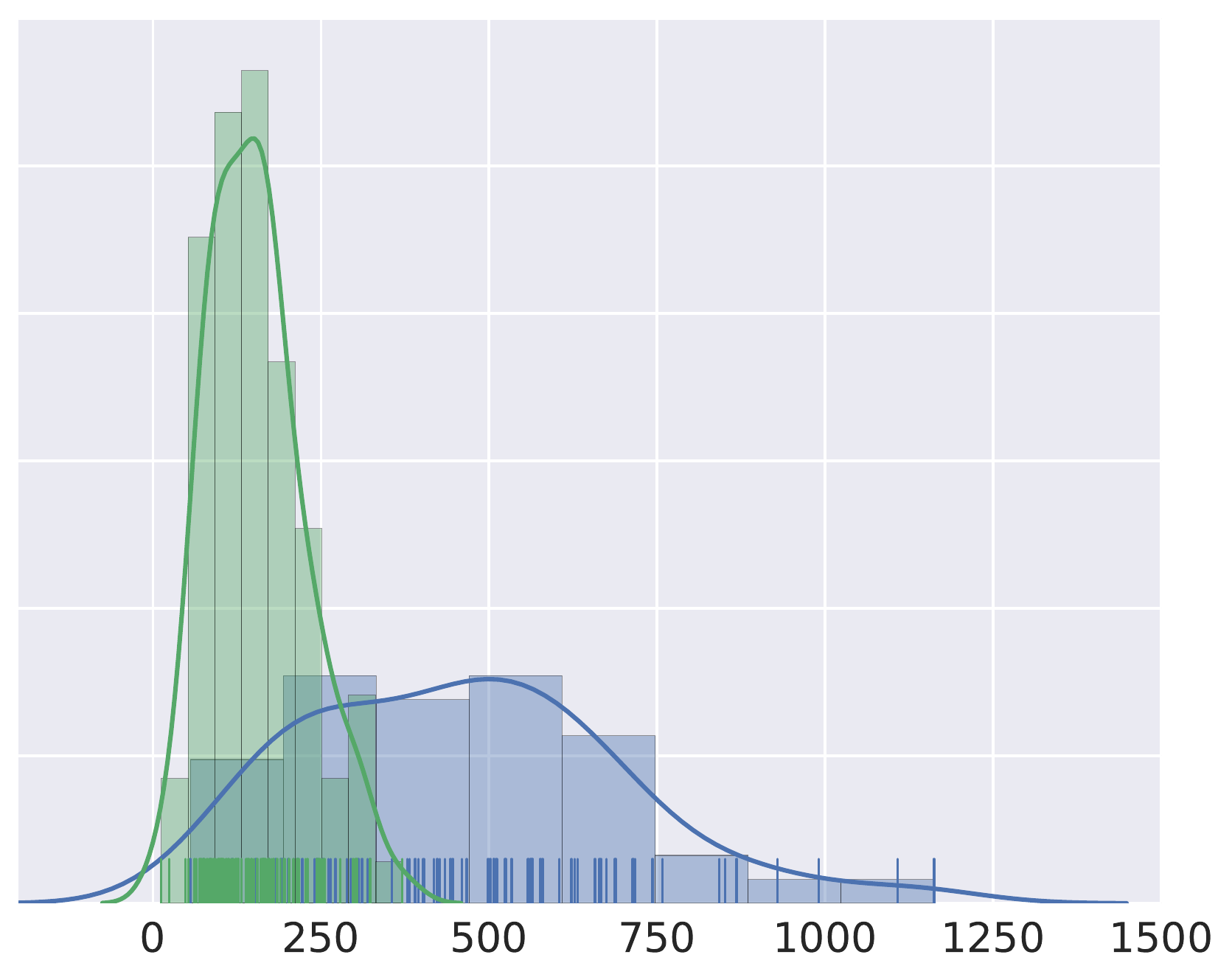}
	\caption[12M-1C distribution plot]{12M-1C Distribution}
	\label{fig:exp:12m-1c-distribution}
  \end{subfigure}\\
  \begin{subfigure}[b]{0.45\columnwidth}
    \includegraphics[width=\textwidth]{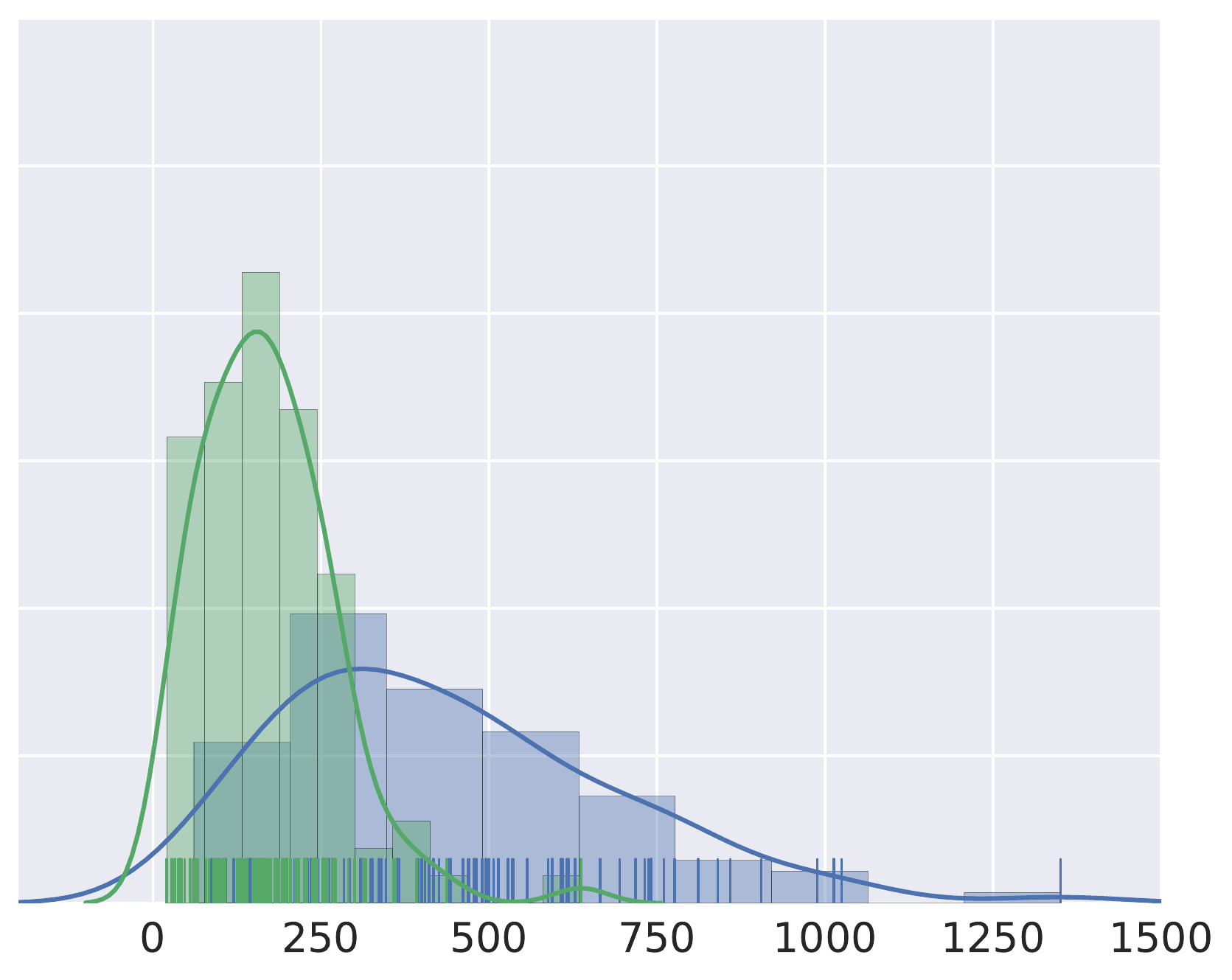}
	\caption[1M-nC distribution plot]{1M-nC Distribution}
	\label{fig:exp:1m-nc-distribution}
  \end{subfigure}
  \begin{subfigure}[b]{0.45\columnwidth}
    \includegraphics[width=\textwidth]{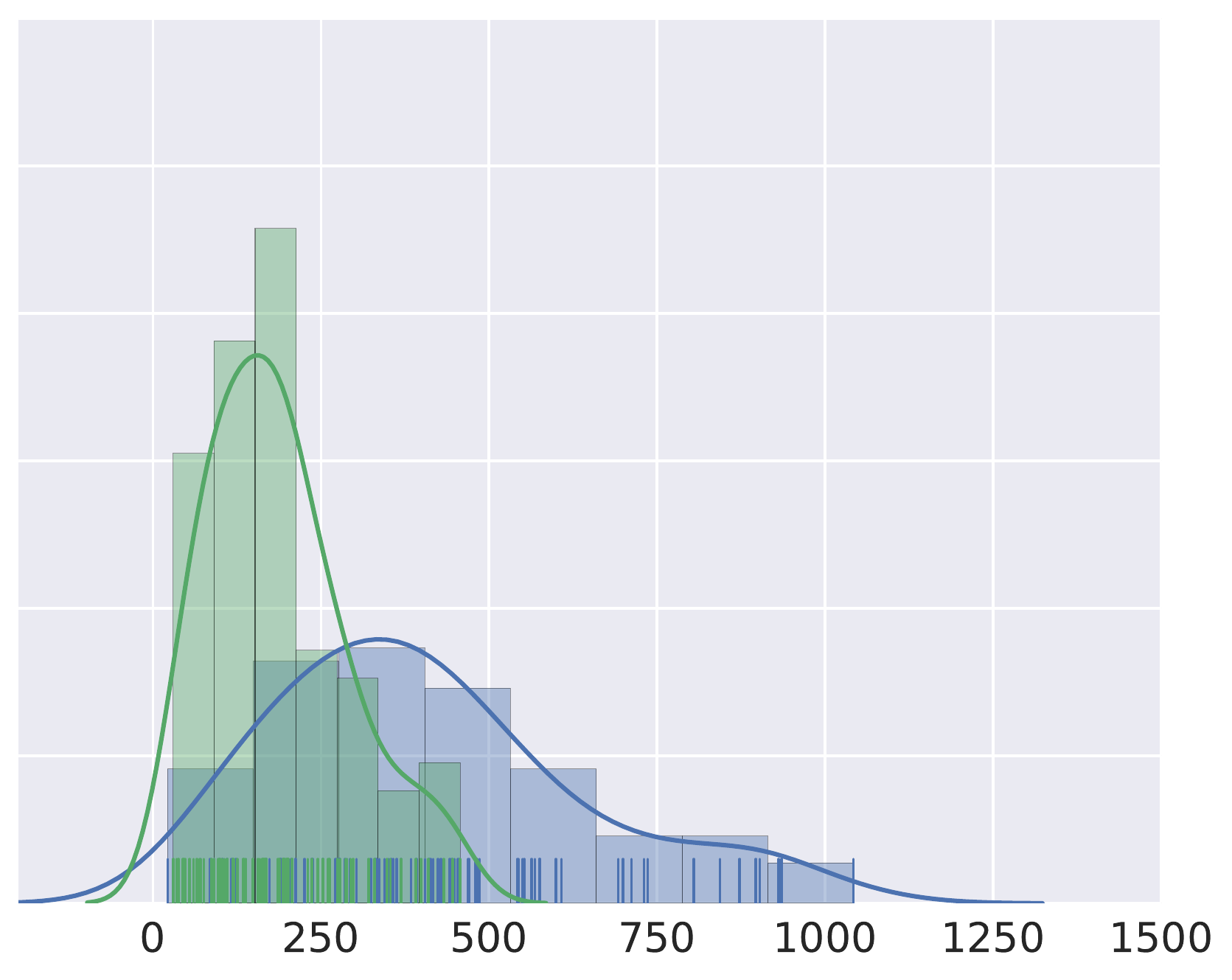}
	\caption[12M-nC distribution plot]{12M-nC Distribution}
	\label{fig:exp:12m-nc-distribution}
  \end{subfigure}
\end{center}
  \caption{Auction results for \acro{quest} across the four conditions: 1M-1C, 12M-1C, 1M-nC and 12M-nC. For each condition, we present both a box plot of travel times, and a representation of the distribution. For both box plot and distribution, blue indicates simulated historical travel times (\acro{hist}), and green indicates the estimated travel time of the auction winner (\acro{auct}).}
  \label{fig:response-times}
\end{figure*}

The results of auction-based allocation under the four experimental conditions
are shown in Tables
\ref{tab:response-times}--\ref{tab:alternate-vehicles}
and Figure \ref{fig:response-times}. Focusing on results obtained using the
\acro{quest} routing engine, under the {\sf 1M-1C} condition, average response times
were reduced from 396 to 205 seconds;
under the {\sf 12M-1C} condition from 460 to 155 seconds; under the {\sf 1M-nC}
condition from 437 to 170 seconds; and under the {\sf 12M-nC} condition from
407 to 187 seconds.

\begin{table}
\begin{center}
\begingroup
\renewcommand*{\arraystretch}{1.1}
\begin{tabularx}{\columnwidth}{p{2.7cm}p{3.5cm}rr}
\toprule
\mbox{Exp.~condition} &		& \acro{gmaps}				& \acro{quest} \\
\midrule
{\sf 1M-1C} & \acro{hist}				& 682.81				& 396.27 \\
& \acro{auct}					& 263.36				& \textbf{205.41} \\
& $t$-statistic ($p$-value)		& 10.5 (\num{8.7e-21}) 	& \qquad 9.7 (\num{1.9e-18}) \\

\midrule
{\sf 12M-1C} & \acro{hist}			& 807.37				& 460.04 \\
& \acro{auct}					& 265.38				& \textbf{154.50} \\
& $t$-statistic ($p$-value)		& 7.99 (\num{1.1e-13}) 	& \qquad 12.28 (\num{9.8e-26}) \\

\midrule
{\sf 1M-nC} & \acro{hist}				& 730.03				& 437.44 \\
& \acro{auct}					& 272.83				& \textbf{170.28} \\
& $t$-statistic ($p$-value)		& 5.29 (\num{3.3e-7}) 	& \qquad 9.02 (\num{1.9e-16}) \\

\midrule
{\sf 12M-nC} & \acro{hist}			& 741.97				& 407.45 \\
& \acro{auct}					& 279.92				& \textbf{186.69} \\
& $t$-statistic ($p$-value)		& 3.51 (\num{5.5e-4}) 	& \qquad 4.83 (\num{3.0e-6}) \\

\bottomrule
\end{tabularx}
\endgroup
\caption{Historical and auction-based response times compared. Values are average response times in seconds with a 2-tailed $t$-statistic.}
\label{tab:response-times}
\end{center}
\end{table}

\begin{table}[t!]
\begin{center}
\begingroup
\renewcommand*{\arraystretch}{1.5}
\begin{tabularx}{.6\columnwidth}{>{\centering\arraybackslash}p{3.7cm}cc}
\toprule
Experimental Condition		& \acro{gmaps} 	& \acro{quest} \\
\midrule
{\sf 1M-1C}  						& 93\%				& 89\% \\
{\sf 12M-1C}  						& 96\%				& 97\% \\
{\sf 1M-nC}  						& 91\%				& 92\% \\
{\sf 12M-nC}  						& 92\%				& 94\% \\
\bottomrule
\end{tabularx}
\endgroup
\caption{Percentage (\%) of times that auction-based allocation chose a different vehicle to that which was dispatched historically.
}
\label{tab:alternate-vehicles}
\end{center}
\end{table}

Table \ref{tab:alternate-vehicles} shows the proportion (percentage) of times
auction-based allocation chose a vehicle to dispatch that was different to the
vehicle that was historically dispatched to an incident. With both the
\acro{quest} and \acro{gmaps} routing engines, a different vehicle was chosen
89+\% of the time across all experimental conditions. This indicates that,
historically, there was often an alternative vehicle that could have reached
an incident location sooner given our assumptions about idle vehicle locations
(discussed below). Note that there is no value judgement inherently attached to the percentage difference values, but it is interesting to be able to consider that a high percentage of differences implies that the methodology evaluated here is predicted to behave differently from the current system, as borne out in the improved response times.

As a whole, these results indicate a potentially large reduction in response
times when using an auction-based approach to dispatching. The auction mechanism
in these experiments produced allocations in a way similar to the ``closest
available vehicle'' strategy currently employed by dispatchers at the LAS, but using a different method to assess what ``closest'' means. 
One factor that may explain the difference in average response times is the accuracy
of the \acro{quest} routing engine when compared with the routing engine used by
the LAS at the time that the data set was recorded. 
The LAS estimate vehicle
travel times using a method that considers the types of road segments along
a proposed route (i.e., number of lanes) but not current or historical traffic
conditions. 
The \acro{quest} routing engine considers historical traffic
conditions and right-of-way rules that apply to emergency services vehicles
when estimating travel times, and so produces more accurate estimates than
state-of-the-art non-specialised alternatives such as \acro{gmaps}.

These results are based on several assumptions. The identities and locations of
idle vehicles were not present in the data set provided by the LAS and needed to
be estimated. Incidents were assumed to be independent: the effect of assigning
a vehicle to an incident, possibly moving it away from responding to subsequent
incidents in its idle area of coverage, were not modelled. Nevertheless, the
auction-based model is attractive because the bid each vehicle agent computes
can be extended to consider factors other than estimated distance or travel
time, factors such as the cost of removing a vehicle from an area of service
(decreasing the equity of coverage), crew fatigue, the ability of a vehicle to
convey a patient, or the presence of specialist equipment or skills of personnel
on board the vehicle. A key feature remains the ability of a routing engine to
accurately estimate travel times, possibly enhanced by real-time traffic data.
These factors that comprise the suitability of a vehicle to respond to an
incident can be clearly presented to a human dispatcher who ultimately makes the
assignment decision.

\section{Related work}
\label{sec:related}

There are four main areas of related work on ambulance dispatch: 
applying new information technologies~(IT);
predicting demand; predicting response time; and identifying the optimum location of
emergency services.

\emph{Applying new technologies} to support emergency response includes a wide range of data-centric modelling and decision-support solutions.
Zhou \emph{et al.}~\cite{zhou-et-al-jasa:2015} created a geo-temporal model of ambulance demand in Toronto (Canada) and demonstrated that such modelling could lead to more accurate predictions of operational results than current industry standards.
This is one of a number of studies that have investigated application of various modelling methodologies to better understand the range of factors that influence emergency response~\cite{penchord,chalk-et-al:2016,dietze-et-al:2003,alcohol-ias:2015}.
A number of approaches for decision-support systems to aid emergency services have been explored, primarily by analysing data from past incidents~\cite{bartels-iscram:2014,barthe-delano-et-al-iscram:2014,sutton-et-al-iscram:2014}.
The problem of providing information to citizens and responders during incidents has been studied by \cite{ruiz-zafra-et-al:iscram:2014}, who focussed in particular on ways to communicate with citizens via mobile devices to provide live updates and instructions.
Zadorozhny \& Lewis~\cite{zadorozhny-lewis:2013} consider the problem of \emph{information fusion}.
Although their example scenario concerns robot-aided urban search-and-rescue, they address the question of data reliability and propose a \emph{crowdsourcing} approach to mitigate the adverse effects of inaccurate or incomplete information, an approach also taken in~\cite{calderon-et-al-iscram:2014}.
Collectively, these studies demonstrate the potential of non-traditional data-backed, technology-based methods to improve ambulance response.

\emph{Predicting incident demand} is perhaps the largest area of EMS
research, and focuses on predicting demand of a population across a day, week
or year. It is important for EMS personnel to understand demand in order
to have appropriate numbers of ambulances on shift.
The moving average method, which is commonly used by ambulance services in the US
to predict demand~\cite{setzler2009ems}, is based on an average of the call volume of one hour time
periods on a specific day in four consecutive weeks over the previous five years.
This can be used to predict demand for a specific location as well as for an entire
city.

Separate models for both daily and hourly demand have been developed by \cite{channouf2007application} based on data for Calgary (Canada) during 2000--2004.
This suggests that there is an overall increase in demand over the four years,
with larger volume in July and December. Special Days, where the demand is
unusually large, can be identified---these include New Year's Day and the
annual Calgary Stampede event. Call arrival data from Toronto is the basis of
the Poisson-based model developed in \cite{matteson2011forecasting}. Here New
Year's Eve and New Year's Day were Special Days. 
Vile \emph{et al.}~\cite{vile2012predicting}
analysed demand data from the Welsh Ambulance Service Trust (UK), once again showing
that there are daily, weekly and yearly periodicities as well as Special Days
(in this case, all Special Days were New Year's Day in different years). There
was also an overall positive increase in demand across the 57 months for which
data was available.

As well as understanding when demand is expected to be particularly high or low,
research also investigates the distribution of demand across geographic regions.
For example, Kamenetzky \emph{et al.}~\cite{kamenetzky1982estimating} developed a model to predict demand
across any area of Southwestern Pennsylvania (US), using regression analysis based on
1979 data from 82 ambulance services in the region. Spatio-Temporal analysis
provides more precise demand models by combining the two techniques discussed
above, predicting demand based on the time of day for specific areas of a
population.
Setlzer \emph{et al.}~\cite{setzler2009ems} developed such a model based on Artificial
Neural Networks to improve prediction forecasts for the Charlotte-Mecklenburg
region of North Carolina (US) beyond the accuracy and precision of the MEDIC
model. Other work has to develop more accurate methods of predicting ambulance
demand for more precise areas
\cite{zhou2015predicting,zhou-et-al-jasa:2015,zhou2016predicting} has also
developed models that are significantly more accurate than MEDIC.

As well as simply predicting ambulance demand, further work investigates
developing models which directly \emph{predict ambulance response time}, based
on a prediction of call arrivals. In this line, Scott \emph{et al.}~\cite{scott1978predicting}
developed a probabilistic model that was fitted to a random 28-day sample from
data for Houston, Texas (US) between July 1973 and June 1974. Similarly 
Taylor~\cite{taylor2017spatial} modelled response times for the London Fire Brigade
using survival analysis, and Thornes \emph{et al.}~\cite{thornes2014ambulance} list factors
affecting response times which include: the number of ambulances; congestion in
A \& E (hospital emergency room); and weather and consequent road conditions.

Further considerations in optimising the ambulance services include the
\emph{location of EMS facilities} to enable adequate coverage across the
city. For example, Gendreau \emph{et al.}~\cite{gendreau1997solving} looked for optimum ambulance
locations in Montr\'{e}al (Canada) under a double coverage model, and
Higgins \emph{et al.}~\cite{higgins2013understanding} describe a spatial model to identify
communities most at risk from fires around Merseyside (UK) based on fire station
location.

Poulton \& Roussos~\cite{poulton2013towards} developed a routing engine
and simulation framework to evaluate the performance of ambulance dispatching
and relocation methods. Their relocation model, which seeks to provide
geographic coverage for current and anticipated emergency incidents, improved
on historical response times on a sample of emergency incidents drawn from
the Greater London (UK) area in 2011.

Lujak \& Billhardt showed that auction-based approaches to ambulance
allocation, applied to a sample of emergency incidents drawn from Madrid in
2009, led to reductions in ambulance travel distance and response times as
compared with a first-come-first-served approach \cite{lujak2013coordinating}.
In contrast to the work presented here, their approach used the \acro{gmaps}
routing engine with a combinatorial auction mechanism, the computational costs
of which scale exponentially with the number of tasks and agents
\cite{berhault2003robot} and is unlikely to be able to handle municipal-sized
dispatching problems.

\section{Summary}
\label{sec:summary}

This paper has described our work applying techniques from multi-robot routing to the problem of ambulance dispatch at the London Ambulance Service.
Our results strongly suggest that a combination of accurate route plan estimation and auction-based vehicle selection has the potential to significantly reduce response times---which was the case for all four experimental conditions we evaluated.
Of the four experimental conditions that we examined, the worst average performance across 100 incidents was for the {\sf 1M-1C} condition (which corresponds to January 2016 in the Harringey CCG), where the average simulated response time for the auction mechanism choice was 48\% faster than the average simulated response time for the historically chosen vehicles.
For the {\sf 12M-nC} condition (2016 across the whole of London), the average simulated response time for the auction mechanism choice was 54\% faster than that of the historically chosen vehicle.

The type of auction-based resource allocation mechanism presented here could also be applied to the task of an ambulance crew deciding which hospital to transport a patient to, termed \emph{conveyance} by the LAS.
It is not always sensible to bring a patient to the nearest hospital due to factors such as a patient's need for access to special equipment, services or medical specialists, the proximity to a patient's home for ease of family visits, the location where a patient has previously been treated, the current waiting time at the hospital's emergency room (termed ``A \& E'' in the UK), or the number of available beds in the hospital.
In the case of stroke patients, for example, it has been shown~\cite{james:2016} that minimising time to treatment --- what is known as ``door to needle time'' --- is best achieved not by conveying patients to the nearest hospital, but by taking them to a specialist stroke unit.
A post-response, pre-conveyance auction could take place, where the ambulance is the ``auctioneer'' and the hospitals are the ``bidders'' to address exactly this issue.

\subsubsection*{Acknowledgements}

This work was partially funded by ESRC through the Data Awareness for Sending Help (DASH) grant ES/P011160/1.
We are grateful to our collaborators from the King's College London Policy Institute, especially Dr Alexandra Pollitt.

\bibliographystyle{splncs04}
\bibliography{mrdash}

\end{document}